\documentclass[superscriptaddress,floatfix,twocolumn,showpacs,preprintnumbers,amsmath,amssymb,prl]{revtex4}

\usepackage{graphicx}
\usepackage{dcolumn}
\usepackage{bm}

\begin{document}

\preprint{}

\title{Cascade of magnetic-field-induced quantum phase transitions in a spin $\bm{\frac{1}{2}}$
triangular-lattice antiferromagnet
}

\author{N. A. Fortune}
\affiliation{Department of Physics, Smith College, Northampton, MA 01063, USA}
\author{S. T. Hannahs}
 \affiliation{National High Magnetic Field Laboratory, 1800 E. Paul Dirac Dr., Tallahassee, FL 32310, USA}
\author{Y. Yoshida}
\altaffiliation[Present address: ]{Institute of Applied Physics and Microstructure Research Center, University
of Hamburg, Jungiusstrasse 11, D-20355 Hamburg, Germany.}\affiliation{Department of Physics, University of
Florida, P.O.\ Box 118440, Gainesville, Florida 32611-8440, USA}
\author{T. E. Sherline}
\altaffiliation[Present address: ]{Neutron Scattering Science Division, Oak Ridge National Laboratory, Oak
Ridge, TN 37831, USA.}\affiliation{Department of Physics, University of Florida, P.O.\ Box 118440, Gainesville,
Florida 32611-8440, USA}
\author{T. Ono}
\author{H. Tanaka}
\affiliation{Department of Physics, Tokyo Institute of Technology, Meguro-ku, Tokyo 152-8551, Japan}
\author{Y. Takano}
\affiliation{Department of Physics, University of Florida, P.O.\ Box 118440,
Gainesville, Florida 32611-8440, USA}

\date{\today}

\begin{abstract}
We report magnetocaloric and magnetic-torque evidence that in Cs$_{2}$CuBr$_{4}$ --- a geometrically frustrated
Heisenberg $S=\frac{1}{2}$ triangular-lattice antiferromagnet --- quantum fluctuations stabilize a series of
spin states at simple increasing fractions of the saturation magnetization $M_{s}$.  Only the first of these
states --- at $M=\frac{1}{3}M_{s}$ --- has been theoretically predicted. We discuss how the higher fraction
quantum states might arise and propose model spin arrangements. We argue that the first-order nature of the
transitions into those states
is due to strong lowering of the energies by quantum fluctuations, with implications for the general character
of quantum phase transitions in geometrically frustrated systems.
\end{abstract}


\maketitle

Geometric frustration appears in a wide variety of physical systems \cite{Ong,Heidarian,Greedan}. In a classical
system, this frustration leads to a large number of states of identical energy. Quantum fluctuations can lift
this degeneracy, creating classically unexpected ground states and excitations.

For one of the simplest possible frustrated systems---a Heisenberg antiferromagnet with spins of quantum number
$S=\frac{1}{2}$ arranged on a triangular lattice---theory predicts that quantum fluctuations should stabilize a
novel up-up-down (\emph{uud}) ground state \cite{Chubukov,Miyahara,Alicea}. Because this collinear state
preserves the continuous rotation symmetry of the spin hamiltonian, low-energy excitations are separated from
the ground state by energy gaps, resulting in the ground state of constant magnetization equal to $\frac{1}{3}$
of the saturation magnetization $M_s$ over a finite field range. Experimentally, however, Cs$_2$CuBr$_4$ is the
only  known $S=\frac{1}{2}$ triangular-lattice antiferromagnet in which this up-up-down state occurs
\cite{Ono,OnoJPCM,OnoJPSJ}. The suppression of this quantum stabilized state with increasing in-plane anisotropy
prevents its formation in the isomorphic compound Cs$_2$CuCl$_4$ \cite{Miyahara,Alicea}.

The spin hamiltonian for Cs$_2$CuBr$_4$ is given by
\begin{equation}
\mathcal{H}=J_1\sum_{<i,j>}\overrightarrow{S_i}\cdot\overrightarrow{S_j}+J_2\sum_{<i,k>}\overrightarrow{S_i}\cdot\overrightarrow{S_k},
\end{equation}
where $J_1=11.3$\,K for nearest-neighbor coupling along the $b$ axis and $J_2=8.3$\,K for weaker
nearest-neighbor coupling within the $bc$ plane \cite{Tsujii}.  Not included in the hamiltonian are two small
perturbations expected to be present:  an antiferromagnetic interlayer coupling that causes the spins to order
at  1.4 K in zero field, and an anisotropic superexchange interaction (Dzyaloshinskii-Moriya) that causes the
spins to lie along the plane of the triangular lattice at zero field. The Dzyaloshinskii-Moriya interaction is
also likely responsible for the suppression of the up-up-down transition in fields applied along the $a$ axis
(perpendicular to the triangular lattice) in Cs$_2$CuBr$_4$. In Cs$_2$CuCl$_4$, each of these is about 5\% of
$J_1$ \cite{Coldea}.

Here we report the complete high-field phase diagram of Cs$_2$CuBr$_4$ up to the saturation magnetic field $H_s=
28.5$\,T. The phase diagram was established through a combination of magnetocaloric and magnetic-torque
measurements. In addition to the expected ordered antiferromagnetic phase at $\frac{M}{M_s} = \frac{1}{3}$, we
find a theoretically unexpected cascade of additional ordered antiferromagnetic phases at higher fractions of
$M_s$.

The magnetocaloric experiment employed a miniature sample-in-vacuum calorimeter \cite{Hannahs} inserted into the
mixing chamber of a dilution refrigerator.  Inside the calorimeter, the 5.35\,mg sample was directly mounted on
a 0.5\,mm $\times$ 1\,mm $\times$ 50\,$\mu$m ruthenium-oxide resistance thermometer with a minimum amount of
nail polish.  The sample and thermometer were weakly thermally linked via 25\,$\mu$m diameter phosphor-bronze
wires to a sapphire ring embedded in a 7.0\,mm diameter silver platform serving as the thermal reservoir. These
wires also served as the electrical leads to the sample thermometer and mechanical support for the sample and
thermometer.

When the magnetic field---produced by a 33\,T resistive magnet and applied along the crystallographic $c$ axis
\cite{Morosin} of the sample---is slowly swept up or down, the magnetocaloric effect produces a temperature
difference $\Delta T$ between the sample and the thermal reservoir that depends on the heat capacity $C_H$, the
temperature dependence of the magnetization $(\partial M/\partial T)_H$, the field sweep rate $\dot{H}$, and the
weak link's thermal conductance $\kappa$ \cite{Scheven}:
\begin{equation}
\Delta T = - \frac{T}{\kappa}\left[\left(\frac{\partial M}{\partial T}\right)_H +\frac{C_H}{T} \frac{d(\Delta
T)}{dH} \right]\dot H.
\end{equation}
Reversing the field sweep direction reverses the sign of the temperature difference, thereby revealing the sign
and magnitude of $(\partial M/\partial T)_H$. Transitions between phases appear as deviations from a smoothly
varying $\Delta T$. First-order phase transitions will also reveal the release/absorption of latent heat as the
sample enters/leaves a lower entropy state. At sufficiently low temperatures, there will also be an additional
heat release as a metastable state gives way to the lower energy stable state for both field-sweep directions
through a first-order transition.

\begin{figure}[btp]
\begin{center}\leavevmode
\includegraphics[width=0.8\linewidth]{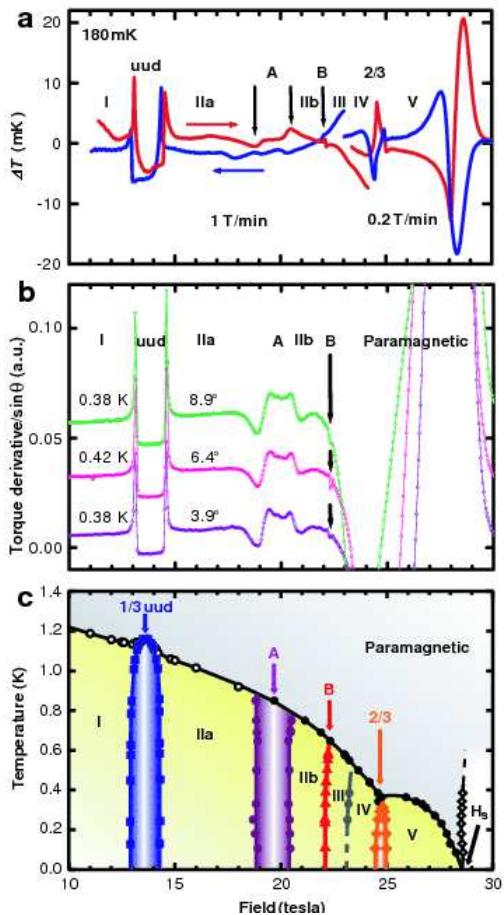}
\caption{(Color online) (a) Evolution of the temperature difference between the sample and thermal reservoir due
to the magnetocaloric effect at 180\,mK, with arrows indicating the field-sweep directions. (b) Derivative of
magnetic torque with respect to $H$ at temperatures near 400\,mK. To produce a torque, the magnetic field was
slightly tilted away from the $c$ axis toward the $b$ axis, by the angle indicated for each curve. (c) Magnetic
phase diagram deduced from the magnetocaloric-effect data taken at various temperatures. Circles indicate
second-order phase boundaries, whereas other symbols except the open diamonds indicate first-order boundaries.
Open diamonds are the positions of the large features near $H_s$ and do not indicate a phase boundary. Lines are
guides to the eye. Data for $H\leq 18$\,T are from Ref.~\cite{Tsujii}, where open circles are from specific
heat. }\label{fig2}\end{center}\end{figure}

Magnetocaloric-effect measurements can be made using swept fields \cite{Scheven}, stepped fields
\cite{Bogenberger}, or modulated fields \cite{McCombe}. The resolution and reproducibility of dc field
magnetocaloric measurements have traditionally been limited by temperature fluctuations, drift, and slow thermal
response, all requiring high sweep rates producing additional heating.  In this experiment, we have overcome
these challenges to swept-field measurements through actively stabilizing the temperature of the thermal
reservoir (sapphire/silver platform), minimizing the heat capacity of the addenda, and reducing the thermal
relaxation time to less than 1 second. The reservoir temperature was maintained  at a constant true temperature
using the algorithm outlined in Ref.~\cite{Fortune} to correct for the magnetoresistance of the sensor.

\begin{figure}[btp]
\begin{center}\leavevmode
\includegraphics[width=0.75\linewidth]{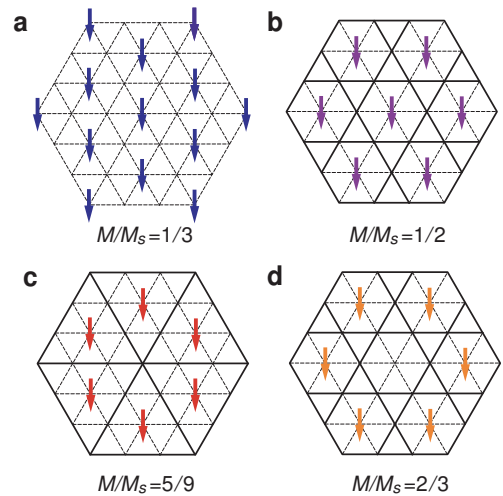}
\caption{(Color online) Collinear states on the triangular lattice at $M/Ms=\frac{1}{3}$, $\frac{1}{2}$,
$\frac{5}{9}$, and $\frac{2}{3}$. Arrows indicate down spins antiparallel to the magnetic field. Vertices with
no arrows indicate up spins pointing in the direction of the field, with broken lines marking rows containing
both spins and solid lines marking rows of only up spins. The A phase may resemble the $M/Ms=\frac{1}{2}$ state,
albeit not collinear.}\label{fig1}\end{center}\end{figure}

Magnetic phase transitions appear as anomalies in the sample temperature as shown in Fig.\ \ref{fig2}a. The
phase diagram deduced from our magnetocaloric-effect data is shown in Fig.\ \ref{fig2}c, along with phase
boundaries for fields $H\leq 18$\,T from Ref.\ \cite{Tsujii}. Additional evidence for this diagram is provided
by the magnetic-torque data shown in Fig.\ \ref{fig2}b. Even for $S=\frac{1}{2}$ spins, theory has long assumed
that the field region above the \emph{uud} phase contains only one coplanar phase \cite{Chubukov}, at least for
the isotropic Heisenberg hamiltonian. We find instead a remarkable cascade of phases in this field region. The
boundaries between these ordered phases are nearly vertical, indicating that the phase diagram is primarily
determined by the zero-temperature energies, not the entropies, of different states. We are witnessing a cascade
of quantum phase transitions.

The \emph{uud} ``plateau" phase appears in the field range 12.9\,T--14.3\,T \cite{Ono,OnoJPCM,OnoJPSJ,Tsujii}.
Below it is phase I, which is known to be incommensurate \cite{OnoJPCM,OnoJPSJ,Fujii}. Above it lies phase IIa,
which is also incommensurate but distinct from phase I \cite{Fujii}. The transitions between the \emph{uud}
phase and phases I and IIa are first-order \cite{OnoJPCM,OnoJPSJ,Tsujii,Fujii2} and the low-lying excitations in
this phase are gapped \cite{Tsujii,Fujii2}.

In the field range 18.8\,T--20.4\,T, a new phase appears, the A phase. As seen in Figs.\ \ref{fig2}a and b, the
transitions to it from phases IIa and IIb are second-order. Phase IIb, in the field range 20.4\,T--22.1\,T, may
in fact be the same as phase IIa.

The magnetization of the A phase corresponds to roughly $\frac{1}{2}$ of the saturation magnetization but forms
no plateau \cite{OnoJPCM}, suggesting that this phase is close to being collinear but is gapless. One likely
arrangement for the nearby collinear state consists of alternating rows of up-down spins and only up spins, as
depicted in Fig.\ \ref{fig1}b, an arrangement predicted to be the $M/M_s=\frac{1}{2}$ ground state of a
triangular-lattice ring-exchange model for two-dimensional solid $^3$He \cite{Momoi}.

The most peculiar of all the new phases is the B phase, appearing at 22.1\,T and only 70\,mT wide. As seen in
Fig.\ \ref{fig2}a, the transitions between the B phase and phases IIb and III are first-order. Like the A phase,
the B phase can be recognized in retrospect as a small feature in the magnetic induction measured in pulsed
magnetic fields \cite{OnoJPCM}. Unlike the A phase feature, however, the feature of the B phase is pointed,
suggesting a magnetization plateau and thus a collinear state with gapped low-lying excitations, at
approximately $\frac{5}{9}$ of $M_s$.

Generalization of Lieb-Schultz-Mattis theorem \cite{Lieb} predicts that any gapped, ordered state must be
commensurate \cite{Oshikawa,Misguich,Hastings}. Indeed, NMR of $^{133}$Cs shows that the B phase is commensurate
\cite{Fujii3}. The collinearity and commensurateness suggest that the B phase may be the $\frac{5}{9}$ state
depicted in Fig.\ \ref{fig1}c, a repetition of two rows of \emph{uud} spins and one row of all up spins. Quantum
calculations of the energy of this $\frac{5}{9}$ state have not yet been performed, but classically, this state
is higher in energy than the coplanar and canted-spiral, three-sublattice states. Therefore, it is most likely that this new collinear, commensurate phase at $\frac{5}{9}$ of $M_s$
--- like the previously known collinear, commensurate phase at $\frac{1}{3}$ of $M_s$ --- owes its existence to
strong quantum fluctuations.

Phase III, in the field region 22.1\,T--23.1\,T, is similar to phases IIa and IIb according to the
magnetocaloric effect, implying that it is also incommensurate. The shapes of the boundaries between the B phase
and phases IIb and III indicate that phase IIb is higher, whereas phase III is lower, in entropy than the B
phase.

Phase IV directly borders on phase III at a second-order transition line. The boundary between this phase and
the high-temperature, paramagnetic phase extrapolates to at most 26\,T at zero temperature, well short of
$H_s=28.5$\,T. This surprising behavior indicates that the ground state of phase IV is higher in energy than a
highly polarized, quantum-mechanically disordered state in the region starting from at least 26\,T and extending
to $H_s$.

The $\frac{2}{3}$-magnetization-plateau phase \cite{OnoJPCM} is observed here in the field region
24.5\,T--25.0\,T. The boundaries between this phase and phases IV and V are first-order. The requirement of
collinearity and commensurateness for ordered magnetization-plateau states implies that the ground state of this
phase should be an arrangement such as shown in Fig.\ \ref{fig1}d. Exact diagonalization for small systems shows
that the ground state at $M/M_s=\frac{2}{3}$ is indeed collinear, for $0.5\lesssim J_2/J_1\lesssim 0.8$
\cite{Miyahara2}. Classically, this $\frac{2}{3}$ state is, like the lower fractional states,
higher in energy than the coplanar and canted-spiral, three-sublattice states.
Stabilization of the commensurate, collinear $\frac{2}{3}$ state observed here appears to imply the existence of
large quantum fluctuations capable of significantly lowering the energy of this state below its classical
expectation.

Phase V covers the highest field region up to $H_s$. In this phase, the magnetization increases steeply with
increasing field \cite{OnoJPCM}, suggesting very rapid suppression of quantum fluctuations by the increasing
field. The shape of the transition line between phase V and the paramagnetic phase is unlike all others in Fig.\
\ref{fig2}c, exhibiting slightly re-entrant behavior at about 25\,T. These two features suggest that the phase
is quite different from phases I--IV. One possibility is that this is a canted spiral phase.

Near $H_s$, the sample temperature exhibits a large peak during an upward field sweep and a deep dip during a
downward sweep (Fig.\ \ref{fig2}a). They indicate a rapid change of entropy with magnetic field, signifying the
emergence of a magnon energy gap at $H_s$.

\begin{figure}[btp]
\begin{center}\leavevmode
\includegraphics[width=0.8\linewidth]{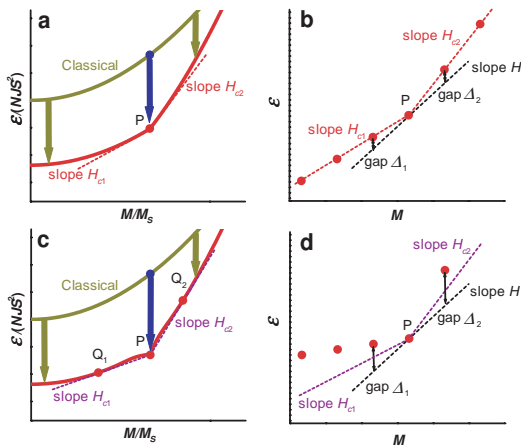}
\caption{(Color online) Ground-state energy $\mathcal{E}$ of a frustrated quantum-mechanical Heisenberg
antiferromagnet as a function of magnetization \emph {M}. (a) Macroscopic behavior of $\mathcal{E}(M)$,
exhibiting a cusp at a gapped ground state, P. (b) Microscopic, extremely expanded view of the region near P,
revealing quantum-mechanically discrete ground states (dots). (c) Macroscopic behavior, when the transitions to
state P are first-order. (d) Corresponding microscopic view. In a and c, the magnetic fields are in
dimensionless units.}\label{fig4}\end{center}\end{figure}

The unexpected cascade of ordered phases within the antiferromagnetic phase boundary of Cs$_2$CuBr$_4$ is quite
unlike the simple phase diagram of the semiclassical, spin-$\frac{5}{2}$, triangular-lattice antiferromagnet
RbFe(MoO$_4$)$_2$ \cite{Smirnov}. The strong contrast demonstrates that even the simplest model of geometrically
frustrated antiferromagnetic interactions is much richer than previously imagined, when it is governed by
quantum mechanics, with important implications for many current theoretical models of superconductivity and
magnetism.

One important implication is, of course, the prospect of new, as yet undiscovered quantum states in such models,
but a second is that the transitions to these states are commonly first order. The transitions to all three
gapped phases observed here---the \emph{uud} phase at $\frac{1}{3}$ $M_{s}$, the very narrow B phase at
$\frac{5}{9}$ $M_{s}$, and the additional plateau phase at $\frac{2}{3}$ $M_{s}$---are first-order. In contrast,
theory usually predicts second-order transitions to a magnetization-plateau-forming state with gapped low-lying
excitations \cite{Chubukov,Damle}. As has been pointed out by Alicea \emph{et al.} \cite{Alicea}, however, one
possible explanation of their first order character could be due to the Dzyaloshinskii-Moriya interaction, which
introduces a cubic term in the free-energy functional.

Here we suggest a new, alternative scenario for consideration. In general, as illustrated in Fig.\ \ref{fig4}a,
when quantum fluctuations select state P as a ground state with energy gaps to the lowest-energy excitations, a
cusp must appear in the ground-state energy $\mathcal{E}$ as a function of magnetization $M$ \cite{Lhuillier}.
These excitations are in fact the two ground states adjacent to P, as shown in Fig.\ \ref{fig4}b. For
second-order transitions, the critical fields $H_{c1}$ and $H_{c2}$ are the two derivatives $\partial
\mathcal{E}/\partial M$ at P. Over the field range between $H_{c1}$ and $H_{c2}$ (Fig.\ \ref{fig4}b), P remains
the lowest state in the ``total" energy $\mathcal{E}-MH$ \cite{Helmholtz}, manifesting itself as a magnetization
plateau. When $H$ is either at $H_{c1}$ or $H_{c2}$, one of the excitations becomes gapless.
We speculate, however, that preferential lowering of P by quantum fluctuations might
produce inflection points in the vicinity of P, as shown in Fig.\ \ref{fig4}c. In that case, the lowest
total-energy state will change discontinuously at $H_{c1}$ and $H_{c2}$ from P to Q$_1$ and Q$_2$ (defined in
Fig.\ \ref{fig4}c). The transitions are now first-order, and are accompanied by non-vanishing energy gaps, as
depicted in Fig.\ \ref{fig4}d. Because this scenario, if verified, relies only on the presence of quantum
fluctuations and not the particulars of the spin-orbit interactions in Cs$_2$CuBr$_4$, it would be applicable to
a broad range of quantum phase transitions in geometrically frustrated systems.

\begin{acknowledgments}
We thank A.\ Wilson-Muenchow, T.\ P.\ Murphy, J.-H.\ Park, and G.\ E.\ Jones for assistance, and J.\ Alicea, Y.\
Fujii, S.\ Miyahara, and O.\ Starykh for discussions. This research was supported by an award from Research
Corporation, a Grant-in-Aid for Scientific Research from the JSPS, the Global CEO Program `Nanoscience and
Quantum Physics' at Tokyo Tech funded by Monkasho, and the National High Magnetic Field Laboratory (NHMFL) UCGP
program and travel grant. The experiments were performed at the NHMFL, which is supported by NSF and the State
of Florida. Y.Y. is a JSPS postdoctoral fellow.
\end{acknowledgments}

\end{document}